\documentclass{article}
\usepackage[T1]{fontenc} 
\usepackage[utf8]{inputenc} 
\usepackage{ismir,amsmath,cite,url}
\usepackage{graphicx}
\usepackage{color}
\usepackage{mathrsfs}
\usepackage{amsfonts}
\usepackage{enumitem}
\usepackage{subcaption}
\usepackage{float}
\usepackage{titlesec}
\usepackage[bookmarks=false,hidelinks]{hyperref}
\usepackage{lineno}
\titlespacing*{\section}{0pt}{2ex}{0.5ex}
\titlespacing*{\subsection}{0pt}{1ex}{0.5ex}

\title{Learning interpretable representation for controllable polyphonic music generation}

\multauthor
{Ziyu Wang \hspace{1cm} Dingsu Wang \hspace{1cm} Yixiao Zhang \hspace{1cm} Gus Xia} {
 \\Music X Lab, Computer Science Department, NYU Shanghai\\
{\tt\small \{ziyu.wang, dingsu.wang, yixiao.zhang, gxia\}@nyu.edu}
}
\def\authorname{Ziyu Wang, Dingsu Wang, Yixiao Zhang, Gus Xia}

\usepackage{subfiles} 

\begin{document}

\maketitle
\begin{abstract}
While deep generative models have become the leading methods for algorithmic composition, it remains a challenging problem to \textit{control} the generation process because the latent variables of most deep-learning models lack good interpretability. Inspired by the content-style disentanglement idea, we design a novel architecture, under the VAE framework, that effectively learns two interpretable latent factors of polyphonic music: chord and texture. The current model focuses on learning 8-beat long piano composition segments. We show that such chord-texture disentanglement provides a controllable generation pathway leading to a wide spectrum of applications, including compositional style transfer, texture variation, and accompaniment arrangement. Both objective and subjective evaluations show that our method achieves a successful disentanglement and high quality controlled music generation.\!\!\footnote{Code and demos can be accessed via \url{https://github.com/ZZWaang/polyphonic-chord-texture-disentanglement}}
\end{abstract}
\section{Introduction}\label{sec:introduction}


With the development of artificial neural networks, deep learning has become one of the most popular techniques for automated music generation. In particular, we see recurrent and attention-based models being able to generate creative and human-like music without heavily handcrafted rules \cite{chen2019effect, musictransformer, poptfm}. 
However, the main drawback of these deep generative models is that they behave like ``black boxes”, and it is difficult to interpret the musical meaning of their internal latent variables \cite{briot2020deep}. Consequently, it remains a challenging task to control the generation process (i.e., to guide the music flow by manipulating the high-level compositional factors such as melody contour, accompaniment texture, style, etc.). This limitation restricts the application scenario of the powerful deep generative models. 

In this paper, we improve the model interpretability for music generation via constrained representation learning. Inspired by the content-style disentanglement idea \cite{music_style_transfer}, we enforce the model to learn two fundamental factors of polyphonic music: \textit{chord} (content) and \textit{texture} (style). The former refers to the representation of the underlying chord progression, and the latter includes chord arrangement, rhythmic pattern, and melody contour. The current design focuses on learning 8-beat long piano composition segments under a variational autoencoder (VAE) framework. 

The core of the model design lies in the encoder. We incorporate the encoder with two \textbf{inductive biases} for a successful \textbf{chord-texture disentanglement}. The former applies a rule-based chord recognizer and embeds the information into the first half of the latent representation. The latter regards music as 2-D images and uses a chord-invariant convolutional network to extract the texture information, storing it into the second half of the latent representation. As for the decoder, we adopt the design from PianoTree VAE \cite{PianoTree}, an architecture that can reconstruct polyphonic music from the latent representation in a hierarchical manner.

We further show that the interpretable representations are \textbf{general-purpose}, empowering a wide spectrum of controllable music generation. In this study, we explore the following three scenarios:
\begin{itemize}[leftmargin=*, itemsep=0pt, parsep=1ex]
    \item[] \textbf{Task 1: Compositional style transfer} by swapping the chord and texture factors of different pieces of music, which can help us re-harmonize or re-arrange a music piece following the style of another piece.
    \item[] \textbf{Task 2: Texture variation} by sampling the texture factor while keeping the chord factor, which is analogous to the creation of ``Theme and Variations” form of composition.
    \item[] \textbf{Task 3: Accompaniment arrangement} by predicting the texture factor given the melody using a downstream encoder-decoder generative model. 
\end{itemize}

In sum, the contributions of our paper are as follows:
\begin{itemize}[leftmargin=*, itemsep=0pt, parsep=1ex,topsep=0pt]
    \item We design a representation disentanglement method for polyphonic music, which learns two interpretable factors: chord and texture.
    
    \item We show that the interpretable factors are general-purpose features for controllable music generation, which reduces the necessity to design heavily-engineered control-specific model architectures. As far as we know, this is the first attempt to explicitly control the compositional texture feature for symbolic polyphonic music generation.
    
     \item We demonstrate that control methods are effective and the quality of generated music is high. Some style transferred pieces are rated even higher than the original ones composed by humans. 
\end{itemize}

\section{Related Work}\label{sec:2}
We review two techniques of automated music generation related to our paper: controlled generation (in Section~\ref{sec:2.1}) and representation disentanglement (in Section~\ref{sec:2.2}). For a more general review of deep music generation, we refer readers to \cite{briot2017, briot2020}.

\subsection{Controlled Music Generation}\label{sec:2.1}
Most existing learning-based methods regard controlled music generation a \textit{conditional estimation} problem. That is, to model $p(\text{music}|\text{control})$, in which both music and control are usually time-series features. Another approach that is closely related to conditional estimation is to first learn the joint distribution $p(\text{music}, \text{control})$ and later on \textit{force} the value of control during the generation process.

The above two methods have been used in various tasks, including generating chords based on the melody \cite{mysong}, creating the melody based on the chords \cite{midinet, ke}, completing the counterparts or accompaniment based on the melody or chord \cite{deepbach, xiaoice, musegan, lakhnes, multivae, poptfm}, and producing the audio waveform based on timbre features \cite{timbretron, tacotron2}.

However, many abstract music factors, such as texture and melody contour, could hardly be explicitly coded by labels. Even if such labels are provided, the control still does not allow continuous manipulation, such as sampling and interpolation. Consequently, it remains a challenging task to control music by more abstract factors without complex heuristics \cite{imposing}.

\subsection{Music Representation Disentanglement}\label{sec:2.2}
Learning disentangled representations is an ideal solution to the problem above, since: 1) representation learning embeds discrete music and control sequences into a continuous latent space, and 2) disentanglement techniques can further decompose the latent space into interpretable subparts that correspond to abstract music factors.  Recent studies show that VAEs \cite{vae, musicvae} are in general an effective framework to learn the representations of discrete music sequences, and the key to a successful disentanglement is to incorporate proper inductive biases into the representation learning models \cite{challenging}. 

Under a VAE framework, an inductive bias can be realized in various forms, including constraining the encoder \cite{f13, wu2020semi, akama2019controlling}, constraining the decoder \cite{drummernet}, imposing multitask loss functions \cite{ec2vae, midivae}, and enforcing transformation invariant results during the learning process \cite{complexbasis, disentangle}. This study is based on our previous work Deep Music Analogy \cite{ec2vae} in which we disentangle pitch and rhythm factors for monophonic segments. We extend this idea to polyphonic composition while the model design is more similar to \cite{wu2020semi}. 

\section{Model}\label{sec:3}

In this section, we introduce the model design and data representation in detail. The goal is to learn the representations of 8-beat long piano compositions (with $\frac{1}{4}$ beat as the shortest unit) and disentangle the representations into two interpretable factors: chord and texture. 


\begin{figure}[ht]
 \centerline{
 \includegraphics[width=\columnwidth]{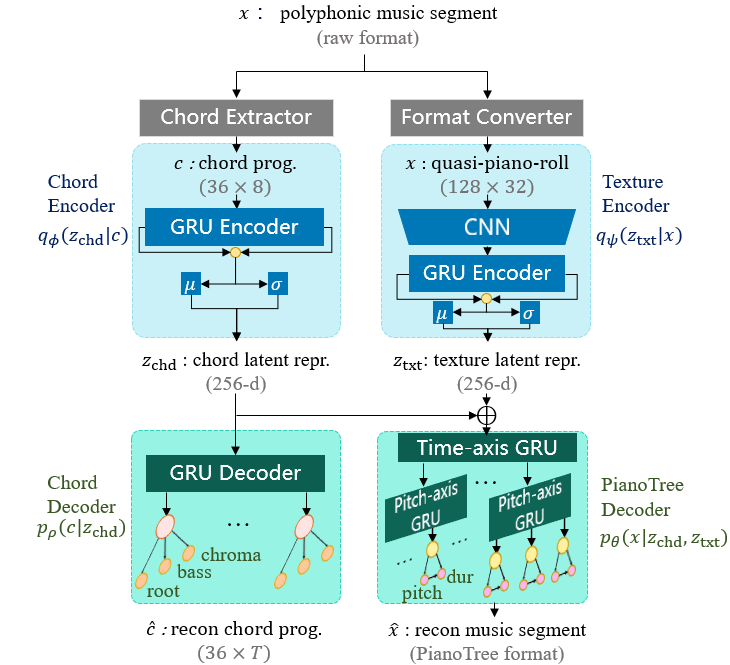}}
 \caption{The model diagram.}
 \label{fig:overall}
\end{figure}

\textbf{Figure 1} shows the overall architecture of the model. It adopts a VAE framework and contains four parts: 1) a chord encoder, 2) a chord decoder, 3) a texture encoder, and 4) a PianoTree decoder. The chord encoder and chord decoder can be seen as a standalone VAE which extracts the latent chord representation $z_{\text{chd}}$. On the other hand, the texture encoder aims to extract the texture representation  $z_{\text{txt}}$ using a chord-invariant convolutional mapping. Finally,  the PianoTree decoder takes in both $z_{\text{chd}}$ and $z_{\text{txt}}$ and outputs the original music in a tree-structured data format.

\subsection{Chord Encoder}\label{sec:3:chd_enc}
The chord encoder first applies rule-based methods \cite{chordextract, mireval} to extract the chord progression under one-beat resolution.  Each extracted chord progression is a 36 by 8 matrix, where each column denotes a chord of one beat. Each chord is a 36-D vector consisting of three parts: a 12-D one-hot vector for the pitch class of the \textit{root}, a 12-D one-hot vector for  the \textit{bass}, and  a 12-D multi-hot \textit{chroma} vector. 


The chord progression is then fed into a bi-directional GRU encoder \cite{musicvae}, and the last hidden states on both ends of the GRU are concatenated and used to approximate the posterior distribution of $z_{\text{chd}}$. Following the assumption of a standard VAE, $z_{\text{chd}}$ has a standard Gaussian prior and follows an isotropic Gaussian posterior. 


Note that although the chord progression here is extracted using algorithms, it can also be provided by external labels, in which case the whole model becomes a conditional VAE \cite{cvae}.

\subsection{Chord Decoder}\label{sec:3:chd_dec}
The chord decoder reconstructs the chord progression from $z_{\text{chd}}$ using another bi-directional GRU. The reconstruction loss of a chord progression is computed as a summation of 8 beat-wise chord loss using cross entropy functions \cite{crossentropy}. For each beat, the chord loss is defined as the product of three parts: 1) the root loss, 2) the bass loss, and 3) the chroma loss. The root and bass are both considered 12-way categorical distributions and the chroma is regarded as 12 independent Bernoulli distributions.

\subsection{Texture Encoder}\label{sec:3:txt_enc}
The input of the texture encoder is an 8-beat segment of polyphonic piece represented by an image-like data format slightly modified from the piano-roll \cite{musegan}. Each 8-beat segment is represented by a 128 by 32 matrix, where each row corresponds to a MIDI pitch and each column corresponds to $\frac{1}{4}$ beat. The data entry at $(p, t)$ records the duration of the note if there is a note onset, and zero otherwise.


The texture encoder aims to learn a chord-invariant representation of texture by leveraging both the translation invariance property of convolution and the blurry effect of max-pooling layers \cite{cnn}. 
We use a convolutional layer with kernel size $12 \times 4$ and stride $1 \times 4$, which is followed by a ReLU activation \cite{relu} and max-pooling with kernel size $4 \times 1$ and stride $4 \times 1$. The convolutional layer has one input channel and 10 output channels.
The convolutional layer design aims at extracting a blurry ``concept sketch'' of the polyphonic texture which contains minimum information of the underlying chord. Ideally, when such blurry sketch is combined with specific chord representation, the decoder can identify its concrete pitches in a musical way.



The output of the convolutional layer is then fed into a bi-directional GRU encoder to extract the texture representation $z_{\text{txt}}$, similar to how we encode $z_{\text{chd}}$ introduced in Section~\ref{sec:3:chd_enc}.

\subsection{PianoTree Decoder}\label{sec:3:pt_dec}
The PianoTree decoder takes the concatenation of $z_{\text{chd}}$ and $z_{\text{txt}}$ as input and decodes the music segment using the same decoder structure invented in PianoTree VAE \cite{PianoTree}, a hierarchical model structure for polyphonic representation learning. The decoder works as follows. First, it generates 32 frame-wise hidden states (one for each $\frac{1}{4}$ beat) using a GRU layer. Then, each frame-wise hidden state is further decoded into the embeddings of individual notes using another GRU layer. Finally, the pitch and duration for each note are reconstructed from the note embedding using a fully-connected layer and a GRU layer, respectively. For more detailed derivation and model design, we refer the readers to \cite{PianoTree}.


\subsection{Training Objective}\label{sec:3:obj}
Let $x$ denote the input music piece and $c = f(x)$ denote the chord progression extracted by algorithm $f(\cdot)$.
We assume standard Gaussian priors of $p(z_{\text{chd}})$ and $p(z_{\text{txt}})$, and denote the output posteriors of chord encoder and texture encoder by $q_\phi(z_{\text{chd}}|c)$, $q_\psi(z_{\text{txt}}|x) $, the output distributions of chord decoder and PianoTree decoder by $p_\rho(c|z_{\text{chd}})$ and $p_\theta(x|z_{\text{chd}}, z_{\text{txt}})$. The objective of the model is:



\begin{align}\label{relativity}
&\mathcal{L}(\phi, \psi, \rho, \theta; x) = \notag\\  
&-\mathbb{E}_{\substack{
z_{\text{chd}}\sim q_\phi\\ \notag 
z_{\text{txt}}\sim q_\psi}}
\bigl[\log p_\rho(c|z_{\text{chd}}) + \log p_\theta(x|z_{\text{chd}}, z_{\text{txt}})\bigr] \notag\\ 
& + \mathrm{KL}(q_\phi||p(z_{\text{chd}})) +
\mathrm{KL}(q_\psi||p(z_{\text{txt}}))\text{.} 
\end{align}


\begin{figure*}[ht!]
	\begin{subfigure}{\textwidth}
		\centering
		\includegraphics[width=\textwidth]{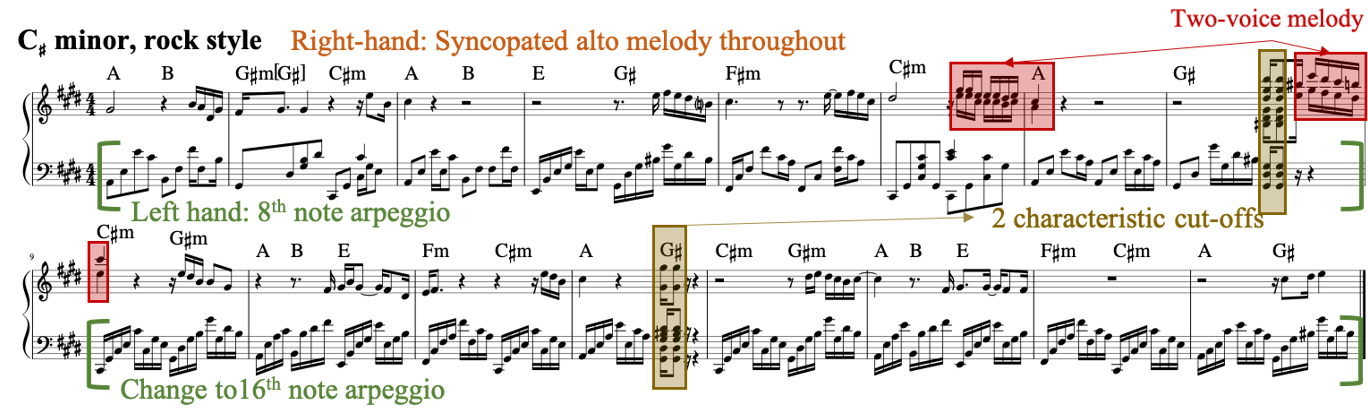}
				\vspace{-1.8\baselineskip}
		\caption{A real piece.}
	\end{subfigure}
	
	\begin{subfigure}{\textwidth}
		\centering
		\includegraphics[width=\textwidth]{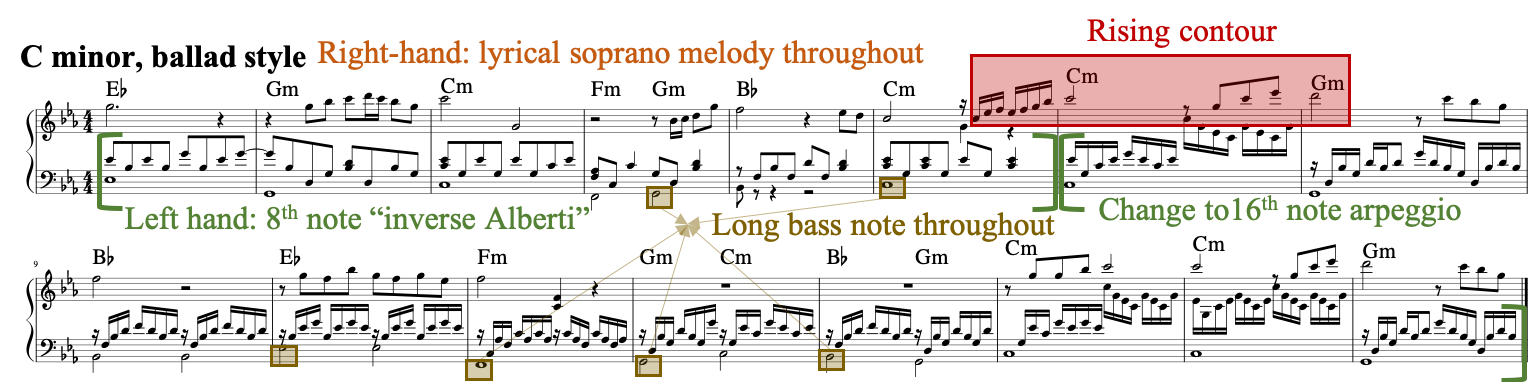}
		\vspace{-1.0\baselineskip}
		\caption{The other real piece.}
	\end{subfigure}
	\begin{subfigure}{\textwidth}
		\centering
		\vspace{-1.0\baselineskip}
		\includegraphics[width=\textwidth]{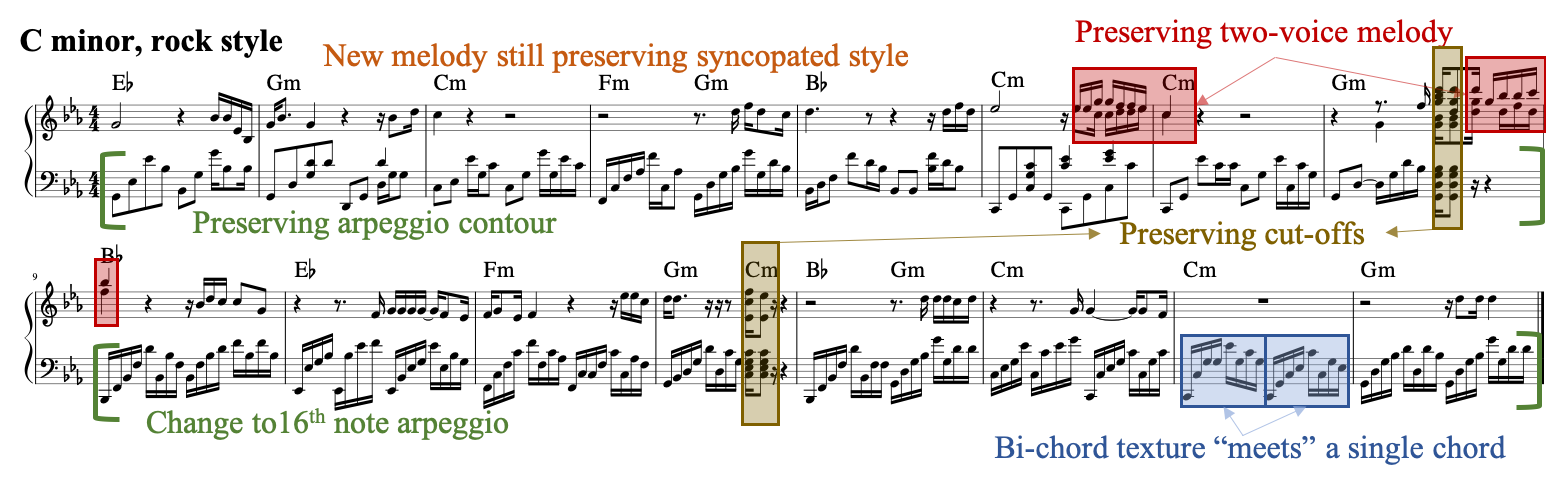}
		\vspace{-1.8\baselineskip}
		\caption{The generated piece by combining $z_{\text{txt}}$ from (a) and $z_{\text{chd}}$ from (b).}
	\end{subfigure}
	\begin{subfigure}{\textwidth}
		\centering
		\vspace{-0.0\baselineskip}
		\includegraphics[width=\textwidth]{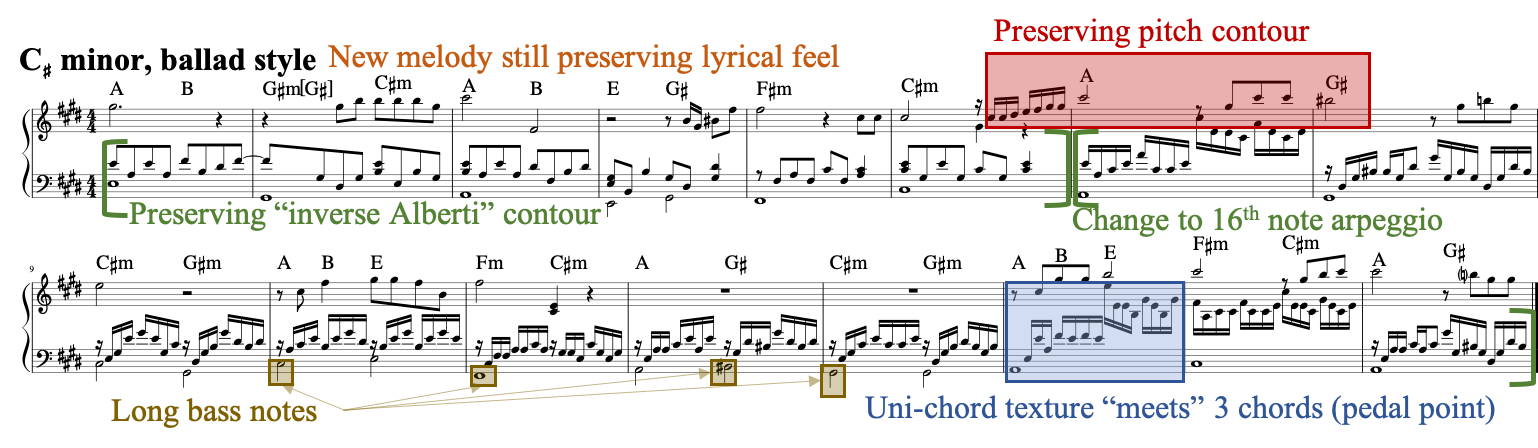}
		\vspace{-1.8\baselineskip}
		\caption{The generated piece by combining $z_{\text{txt}}$ from (b) and $z_{\text{chd}}$ from (a).}
	\end{subfigure}
	\caption{An example of compositional style transfer of 16-bar-long samples when $k = 2$. }
	\label{fig:longswap}
\vspace{-1.0\baselineskip}
\end{figure*}

\section{Controlled Music Generation}\label{sec:4}
In this section, we show some controlled generation examples of the three tasks mentioned in the introduction.

\subsection{Compositional Style Transfer}\label{sec:4:styletransfer}



By regarding chord progression \textit{content} and texture \textit{style}, we can achieve compositional style transfer by swapping the texture representations of different pieces. Figure~\ref{fig:longswap} shows the transferred results ((c) \& (d)) based on two 16-bar samples ((a) \& (b)) in the test set by swapping $z_\text{txt}$ every 2 bars (without overlap).\!\!\footnote{The presented excerpts are converted from MIDI by the authors. The chord labels are inferred from the original/generated samples.}

\begin{figure*}[htpb]
	\begin{subfigure}{\textwidth}
		\centering
		\includegraphics[width=\textwidth]{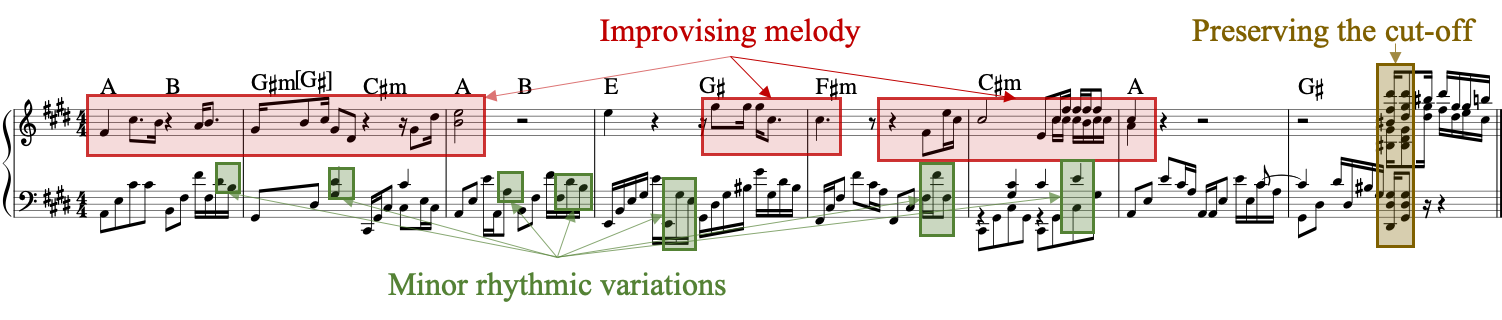}
		\vspace{-1.8\baselineskip}
		\caption{An example of posterior sampling of $z_{\text{txt}}$ of the first 8 bars of the segment (a) in \figref{fig:longswap}}
		\label{fig:post}
	\end{subfigure}
	
	\begin{subfigure}{\textwidth}
		\centering
		\includegraphics[width=\textwidth]{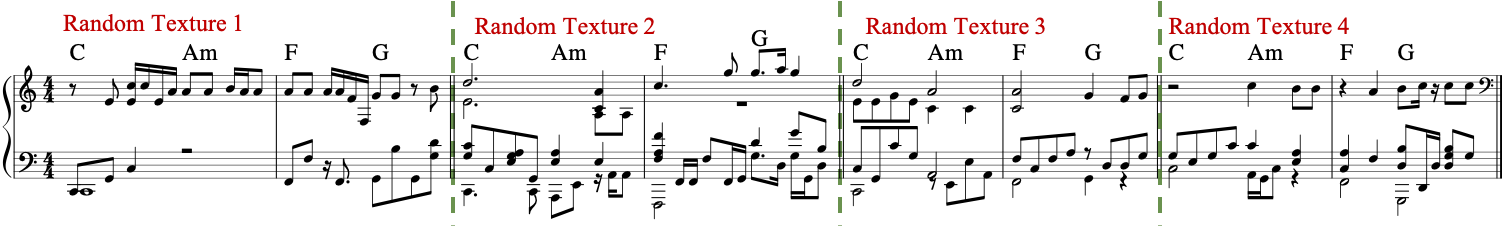}
		\caption{An example of prior sampling of $z_{\text{txt}}$ under given chord progression C-Am-F-G. Each two-bar segment is independently sampled, having different texture.}
		\label{fig:prior}
	\end{subfigure}
	\vspace{-1\baselineskip}
	\caption{Examples of texture variations via posterior sampling and prior sampling.}
	\label{fig:txt_variation}
	\vspace{-1\baselineskip}
\end{figure*}

We see that such long-term style transfer is successful: The generated segment (c) follows the chord progression of (b) while mimicking the texture of (a), while (d) follows the chord progression of (a) while mimicking the texture of (b). As shown in the marked scores, the style transfer is effective. E.g., the cut-offs, melody contours, and the shape of the left-hand accompaniment are all preserved. 


\subsection{Texture Variation by Sampling}\label{sec:4:vartxt}
We can make variations of texture by sampling from $z_{\text{txt}}$ while keeping $z_{\text{chd}}$. Here, we investigate two sampling strategies: sampling from the posterior $q_\psi(z_{\text{txt}}|x)$, and sampling from the prior $p(z_\text{txt})$.


Sampling from the posterior distribution $q_\psi(z_{\text{txt}}|x)$ yields reasonable variations as shown in \figref{fig:post}. The variations of the right-hand melody can be seen as an improvisation following the chord progression and the melody. On the contrary, there is only small variation in the left-hand part, showing that the model regards the left-hand accompaniment as the dominant feature of texture. 

Sampling from the prior distribution $p(z_{\text{txt}})$ changes the texture completely. Figure~\ref{fig:prior} shows a series of examples of prior sampling under the same chord progression C-Am-F-G. The resulting generations follow exactly the chord progression but with new textures. 

\begin{figure*}[ht]
 \centerline{
 \includegraphics[width=\textwidth]{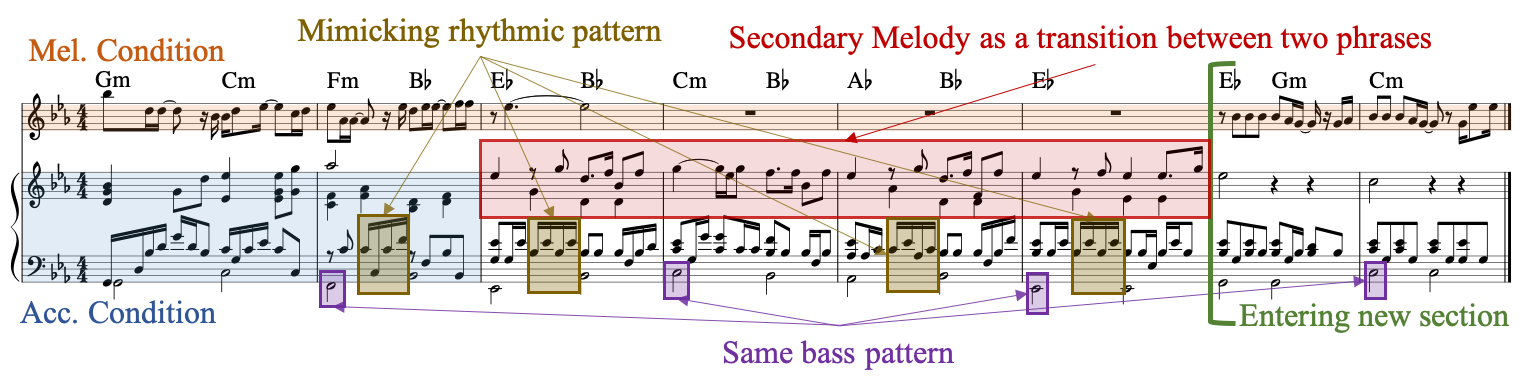}}
  \vspace{-1\baselineskip}
 \caption{An example of accompaniment arrangement conditioned on melody, chord progression, and first 2 bars of accompaniment.}
 \label{fig:arr}
 \vspace{-1\baselineskip}
\end{figure*}

\subsection{Accompaniment Arrangement}\label{sec:4:accarr}
We use a downstream predictive model to achieve accompaniment arrangement. For this task, we provide extra vocal melody tracks paired with the piano samples, and the model learns to generate 16-bar piano accompaniment \textit{conditioned} on melody in a supervised fashion. 

We encode the music every 2 bars (without overlap) into latent representations. For the accompaniment, we use the proposed model to compute the latent chord and texture representation, denoted by $\mathbf{z}_{\text{chd}} = [z_{\text{chd}}^{(1)}, ..., z_{\text{chd}}^{(4)}]$ and $\mathbf{z}_{\text{txt}} = [z_{\text{txt}}^{(1)}, ..., z_{\text{txt}}^{(4)}]$. For the melody, we use the EC$^2$-VAE \cite{ec2vae} to compute the latent pitch and rhythm representations, denoted by $\mathbf{z}_{\text{p}} = [z_{\text{p}}^{(1)}, ..., z_{\text{p}}^{(4)}]$ and $\mathbf{z}_{\text{r}} = [z_{\text{r}}^{(1)}, ..., z_{\text{r}}^{(4)}]$. Then, we adopt a vanilla Transformer \cite{transformer} to model $p( \mathbf{z}_{\text{txt}},\mathbf{z}_{\text{chd}} |\mathbf{z}_{\text{p}}, \mathbf{z}_{\text{r}})$, in which the encoder takes in the condition and the decoder's input is a shifted right version $[\mathbf{z}_{\text{chd}}, \mathbf{z}_{\text{txt}}]$. Both encoder and decoder inputs are incorporated with a \textit{positional encoding} indicating the time positions and a learned \textit{factor embedding} indicating the representation type (i.e., pitch, rhythm, chord or texture). 

\figref{fig:arr} shows an example of accompaniment arrangement, where the first staff shows the melody and the second staff shows the piano accompaniment. In this case, the whole melody, together with the complete chord progression and the first 2 bars of accompaniment are given. The chord conditioning is done by forcing the decoded chord representation to match the given input during inference time. (A similar trick is used in \cite{lakhnes}.) From \figref{fig:arr}, we see that the model predicts a similar texture to the given accompaniment. Moreover, it fills in a secondary melody line as a transition when the lead melody is rest. 

Note that the arrangement can be generated in a flexible way by conditioning on different sets of latent factors. Much longer examples and more conditioning settings are available on our github page.

\section{Experiments}\label{sec:5}
\subsection{Dataset and Training}\label{sec:5:datatrain}
We train our model on the POP909 dataset\cite{pop909}, which contains about 1K MIDI files of pop songs (including paired vocal melody and piano accompaniment). We further extract the chord annotations using \cite{chordextract, mireval}. We only keep the pieces with $\frac{2}{4}$ and $\frac{4}{4}$ meters and cut them into 8-beat music segments (so that each data sample in our experiment contains 32 time steps under 16$^{\text{th}}$ note resolution). In all, we have 66K samples. We randomly split the dataset (at song-level) into training set (90\%) and test set (10\%). All training samples are further augmented by transposing to all 12 keys.

In our experiment, the VAE model uses 256, 512, and 512 hidden dimensions for the GRUs in chord encoder, chord decoder and texture encoder respectively. The latent dimension of $z_{\text{chd}}$ and $z_{\text{txt}}$ are both 256. The model size of the PianoTree decoder is the same as the implementation in the original paper \cite{PianoTree}. The transformer model has the following size: hidden dimension $= 256$, number of layers $=4$ and number of heads $= 8$. 

For both models, we use Adam optimizer \cite{adam} with a scheduled learning rate from 1e-3 to 1e-5. Moreover, for the VAE model, we use KL-annealing \cite{klanealing}, i.e. setting a weight parameter for the KL-divergence loss starting from 0 to 0.1. We set batch size to be 128 and the training converges within 6 epochs. For the downstream transformer model, we use 12K warm-up steps for learning rate update \cite{warmup}. We use the same batch size and the model converges within 40 epochs.

\subsection{Objective Measurement}\label{sec:5:objmeasure}
When $z_{\text{chd}}$ and $z_{\text{txt}}$ are well disentangled, small variations over the note pitches of the original music should lead to a larger change on $z_{\text{chd}}$, while variations of rhythm will influence more on $z_{\text{txt}}$. Following this assumption, we adopt a \textit{disentanglement evaluation via data augmentation} method used in \cite{factorvae} and further developed in \cite{ec2vae}.

\begin{figure}[htpb]
	\begin{subfigure}{\columnwidth}
		\centering
		\includegraphics[width=\textwidth]{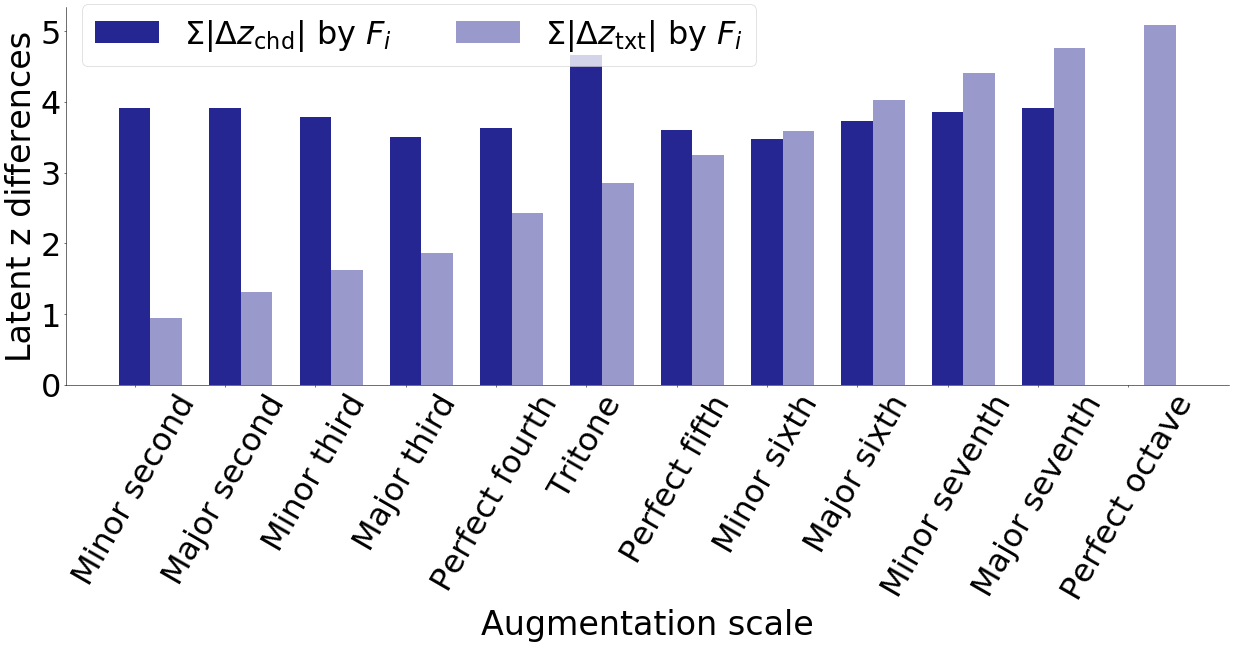}
		\caption{A comparison between $\Delta z_{\text{chd}}$,  $\Delta z_{\text{txt}}$ after pitch transposition on all notes.}
	    \label{fig:exp1}
	\end{subfigure}
	
	\begin{subfigure}{\columnwidth}
		\centering
		\includegraphics[width=\textwidth]{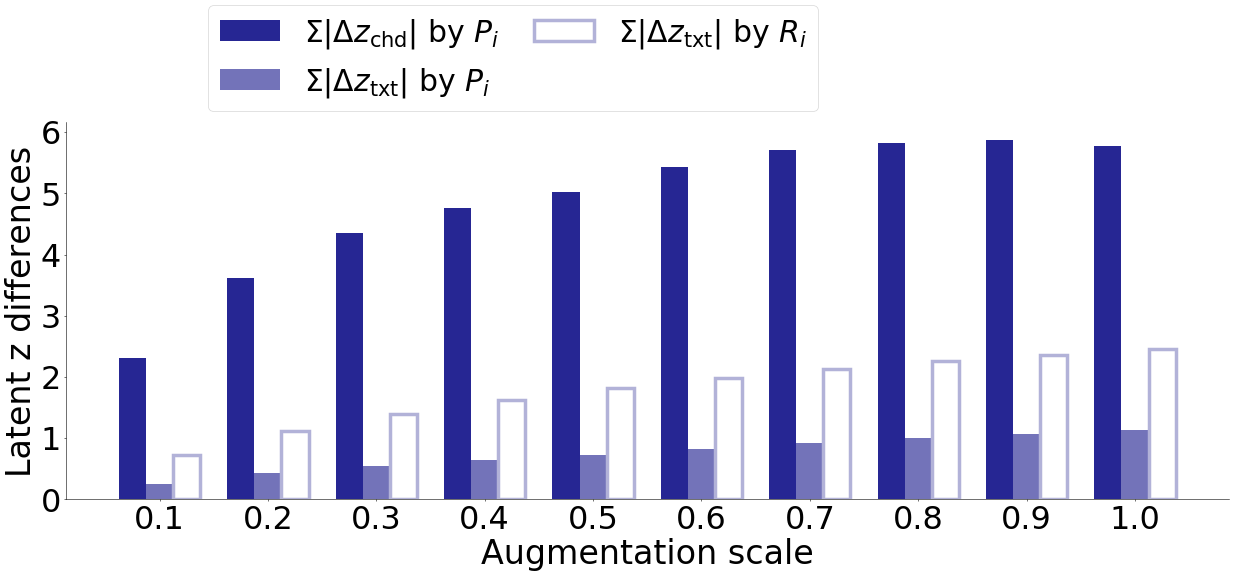}
		\caption{A comparison among $\Delta z_{\text{chd}}$,  $\Delta z_{\text{txt}}$ after beat-wise pitch transposition and texture augmentation with different probabilities.}
	    \label{fig:exp2}
	\end{subfigure}
	\caption{Results of objective measurement.}
	\label{fig:obj-exp}
\end{figure}

 We define $F_i$ as the operation of transposing all the notes by $i$ semitones, and use the $L_1$-norm to measure the change of latent $z$ after augmentation. Figure \ref{fig:exp1} shows a comparison between $\Sigma |\Delta z_{\text{chd}}|$ and $\Sigma |\Delta z_{\text{txt}}|$ when we apply $F_i$ to all the music pieces in the test set (where $i \in [1, 12]$).


It is conspicuous that when augmenting pitch in a small range, the change of $z_{\text{chd}}$ is much larger than the change of $z_{\text{txt}}$. At the same time, the change of $z_{\text{txt}}$ gets higher as the augmentation scale increases. Similar to the result in \cite{ec2vae}, the change of $z_{\text{chd}}$ reflects human pitch perception as $z_{\text{chd}}$ is very sensitive to a tritone transposition, and least sensitive for a perfect octave.

We further define $P_i$ as the function to randomly transpose all the notes in one beat either up or down one semitone under a certain probability $i$, and $R_i$ as the function to randomly reduce the note duration by half. Figure \ref{fig:exp2} shows a comparison between $\Sigma |\Delta z_{\text{chd}}|$ and $\Sigma |\Delta z_{\text{txt}}|$ when we apply $P_i$ and $R_i$ to all the music pieces in our test set (where $i \in [0.1, 1.0]$).


For each value of $i$ in the figure \ref{fig:exp2}, the first and second bars demonstrate $\Sigma |\Delta z_{\text{chd}}|$ and $\Sigma |\Delta z_{\text{txt}}|$ caused by $P_i$ function, while the third bar indicates $\Sigma |\Delta z_{\text{txt}}|$ caused by $R_i$ function. (We did not show $\Sigma |\Delta z_{\text{chd}}|$ caused by $R_i$ since they are all zero.) It again proves that the chord representation is more sensitive than texture representation under pitch variations, and conversely, texture representation is more sensitive than chord representation under rhythm variations. 


\subsection{Subjective Evaluation}\label{sec:5:subeval}

Besides objective measurement, we conduct a survey to evaluate the musical quality of compositional style transfer (see Section~\ref{sec:4:styletransfer}). Each subject listens to ten 2-bar pieces with different chord progressions, each paired with 5 style-transfer versions generated by swapping the texture representation with a random sample from the test set. In other words, each subject evaluates 10 groups of samples, each of which contains 6 versions of textures (1 from the original piece and 5 from other pieces) under the same chord progression. Both the order of groups and the sample order within each group are randomized. After listening to each sample, the subjects rate them based on a 5-point scale from 1 (very low) to 5 (very high) according to three criteria: \textit{creativity}, \textit{naturalness} and \textit{musicality}.
\begin{figure}[htb]
 \centerline{
 \includegraphics[width=\columnwidth]{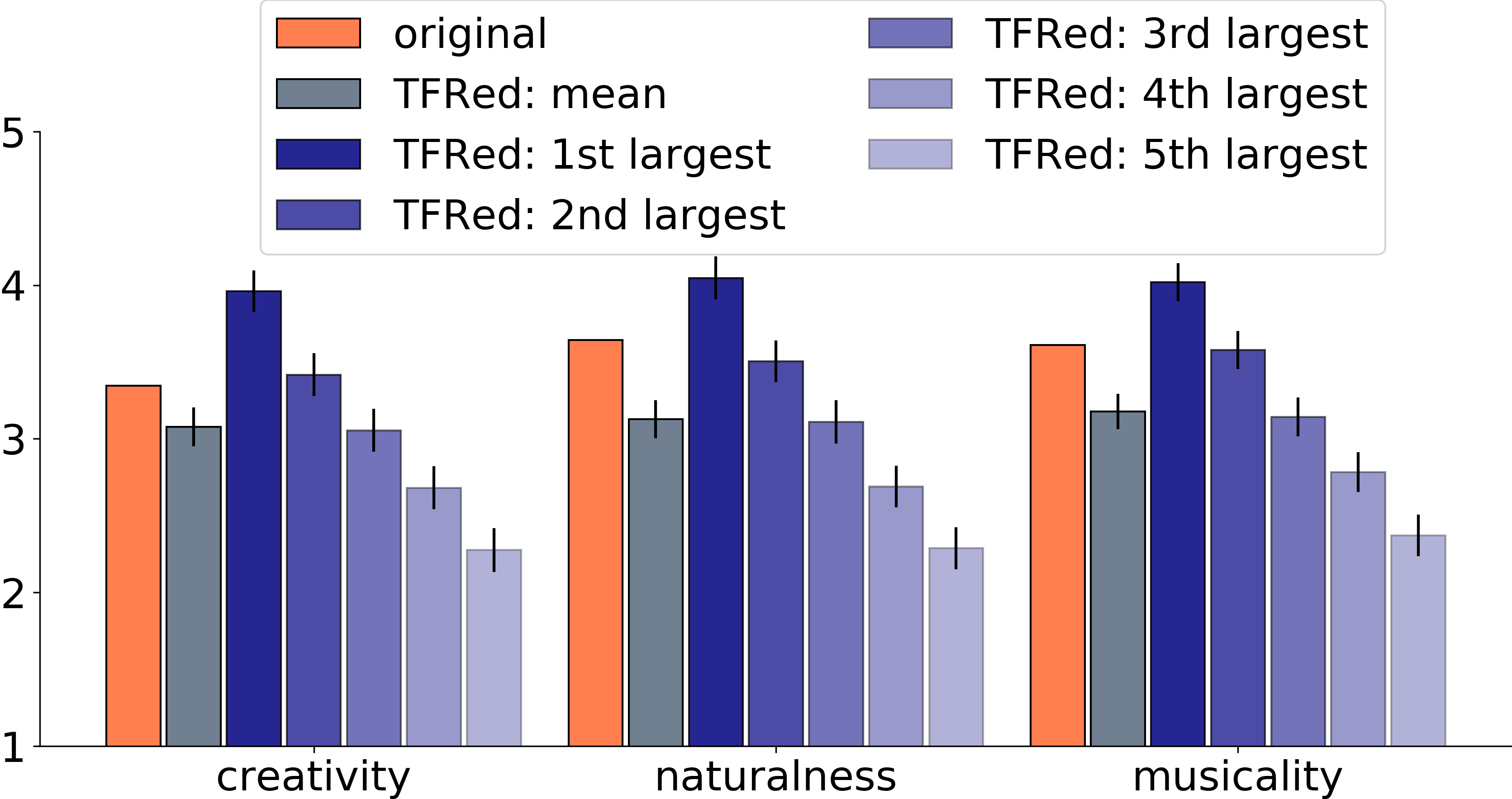}}
 \caption{Subjective evaluation results. Here ``TFRed:~$x$th largest'' denotes the $x^{\text{th}}$ (largest) order statistic of the transferred segments.}
 \label{fig:barplot}
\end{figure}

A total of 36 subjects (26 females and 10 males) participated in the survey. \figref{fig:barplot} shows the comparison result among the original pieces (indicated by the orange bars) and the transferred pieces in terms of their mean and \textit{order} statistics. The heights of bars represent averaged ratings across the subjects and the error bars represent the confidence intervals computed via paired t-test \cite{ttest}. The result shows \textit{if we randomly transfer a piece's texture 5 times, the best result is significantly better than the original version (with p-value $< 0.005$)}, and there are only marginal differences between the second-largest statistics and the original (with p-value $> 0.05$) in terms of creativity and musicality. We also see that on average the transferred results are still rated lower than the original ones. How to automatically decide the quality of a transferred result is considered a future work. 

\section{Conclusion and Future Work}\label{sec:6}
In conclusion, we contributed an effective algorithm to disentangle polyphonic music representation into two interpretable factors, chord and texture, under a VAE framework. Such interpretable representations serve as an intuitive human-computer co-creation interface, by which we can precisely manipulate individual factors to control the flow of the generated music. In this paper, we demonstrated three ways to interact with the model, including compositional style transfer via swapping the latent codes, texture variation by sampling from the latent distribution, accompaniment arrangement using downstream conditional prediction, and there are potentially many more. We hope this work can shed light on the field of controllable algorithmic composition in general, especially on the paradox between model complexity and model interpretability.

We acknowledge that the learned music factors are still very basic. In the future, we plan to extract more abstract and longer-range features using hierarchical models. We also plan to explore more ways to control the music generation for practical usage. 


\bibliography{ISMIRtemplate}

%
%
%
%

\end{document}